# Exploring System Resiliency and Supporting Design Methods

**James J. Cusick, PMP**
IEEE Computer Society Member
New York, NY
j.cusick@computer.org

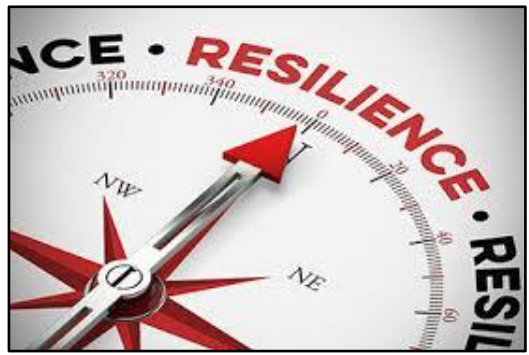

## Abstract

*This paper provides a survey of the industry perspective on System Resiliency and Resiliency design approaches and briefly touches on Organizational Resiliency topics. Beginning with a composite definition of Resiliency, System Capabilities, Adversities, and the Resiliency Lifecycle the document then covers Operational Response Timelines, Failure Sources and Classifications. Next, Design for Resiliency is discussed with an introduction to Systems Theory and a review of Tradeoff Analysis and Resiliency Dependencies. Then more than a dozen Resiliency Design Patterns are included for the reader to consider for their own solutioning. Supporting non-functional design topics including Availability, Performance, Security, Reliability as well as Reliability Allocation using Reliability Block Diagrams are also covered. Additionally, Failure Mode and Effect Analysis is reviewed, and a Resiliency Maturity Model is discussed. Finally, several Resiliency Design Examples are presented along with a set of recommendations on how to apply System Resiliency concepts and methods in an IT environment.*

# Table of Contents

# Exploring System Resiliency and Supporting Design Methods

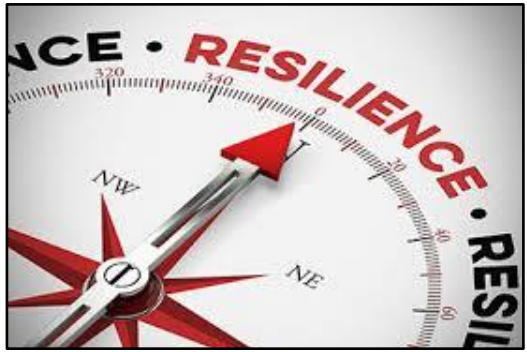

## 1  Overview

We all know what resilience means. It is exemplified by the person who keeps on going no matter the odds and no matter the punishing conditions they face. We can think of countless cultural references in sports or other situations where despite reaching any reasonable point of no return an individual or team keeps going, recovers, pushes through, and completes the game, mission, or task. Call it gumption, endurance, reserve energy, or what have you. But in the end, it is resiliency.

From a systems point of view resilience has a specific meaning which has a formal definition. We will provide this below. In summary, system resilience is the ability to recovery from unplanned events. This paper provides a detailed survey of the industry perspective on resiliency, primarily on  system resiliency and briefly touching on organizational resiliency topics. This document also discusses response patterns to resiliency from engineering challenges. Furthermore, this paper presents the several technical approaches around resiliency with a critical review of these each considering the globally recognized view of resiliency in the industry.

This analysis can provide a foundation around how resiliency should be considered and defined within an IT organization, provide best practices on how resiliency is managed in an organization, and provide a roadmap on how to develop a resiliency strategy for the future.

### 1.1  Scope

This document primary addresses system resiliency. Organizational resiliency is a broader topic which typically covers Business Continuity including the management of people, information, technology, and facilities. This document will provide a framework for understanding many of the core principles which drive decisions in these areas, but the focus will be on technological systems or IT and software systems as they support the a given business. This document focuses on first defining System Resiliency and then providing methods for System Resiliency design.

# 2 System Resiliency Defined

In computing, resiliency means that your infrastructure or software solution can "take a beating and keep on ticking". Essentially, a system is resilient if it continues to carry out its mission in the face of ***adversity*** (i.e., if it provides required ***capabilities*** despite excessive stresses that can cause disruptions). Being resilient is important because no matter how well a system is engineered, reality will sooner or later conspire to disrupt the system (Firesmith, 2019).

Here are two formal definitions of System Resiliency:

- *System resilience is an ability of the system to withstand a major disruption within acceptable degradation parameters and to recover within an acceptable time. (*Hadji-Janev, 2015)

- *[Resilience is] … "the ability of a system to withstand a major disruption within acceptable degradation parameters and to recover within a suitable time and reasonable costs and risks.* (Madni, 2020)

Note that resiliency is not the same thing as being reliable or available or robust or even fault tolerant. These concepts and comparisons will be explored below.

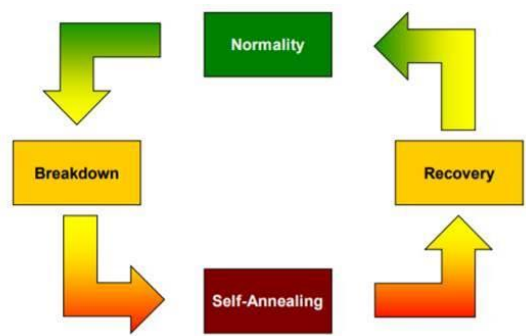

*Figure 1 – Standard computing resiliency model (Urena, 2020)*

An essential process model abstraction for resiliency is shown in Figure 1. This loop is simple. The system moves from normal operations to a disruption, corrects itself preferably **automatically** through an operational degradation period, and then recovers to normal operation. The key point here is that the *system degrades* and *recovers*. It ***does not fail***.

## 2.1 What is a Capability

Just as we might think of applications in terms of features, in a resiliency model, we think of systems in terms of capabilities which need to be present to allow for a system to achieve a resilient behavior profile. Among other capabilities, a resilient system must have at minimum the following (Madni, 2020):

1. **Capacity to rebound**: This implies that a system is able to be restore to its earlier state.
2. **Capacity for Resilience** (as opposed to Robustness):
    a. Robustness is "**dependability** with respect to external faults".

    b. Resilience addresses system "**adaptability** which invariably requires some form of structural change". This also introduces the concept of *graceful degradation.*
3. **Capacity for "adaptive capacity"**: This concept expands the system capabilities in the face of adversity giving us "graceful extensibility".

## 2.2 Defining Adversity

The reason these capabilities are important is that the world is a dangerous place. For software and systems, we often talk about internal design flaws as faults and failures and those are certainly issues that need to be planned for from a design, quality management, and operations perspective. However, from a system engineering perspective there is also the angle of hazard analysis. As an example, if I design an aircraft and it has no intrinsic flaws but I then fly it into a flock of birds that then becomes a clear hazard to the engines and the overall system needs to be designed for that eventuality. This is what we mean by an example of an adversity.

As a further example, not all domains are equal in their level of danger. This then drives the quality needs of the system in order to responds to those adverse effects in appropriate levels.

> *Adversities can be human-made or natural and may originate within the system (endogenous adversity) or from without the system (exogenous adversity). Exogenous [sic] adversities include inclement weather, natural disasters, and adversaries with intelligence and intent. (*Ferris, 2019)

Oftentimes when we consider a typical IT environment we might not think of the systems nature of the domain. However, there are many components at play including external suppliers, vendors, logistical entities, various computing environments, global supply chains, and more. For many companies significant events including technical, cybersecurity, and most recently the COVID-19 pandemic has disrupted operations. These are precisely the types of adversities that a resiliency engineering perspective would account for.

## 2.3 Resiliency Lifecycle

With a set of definitions now in hand it is possible to now introduce a standard resiliency event lifecycle. Like an ITIL Incident Management response lifecycle (Cusick, 2010) where we see a fault generating an outage followed by a repair (and measured by MTTF and MTTR), the resiliency lifecycle is similarly modeled as in Figure 2 below provides a similar event and recovery response cycle. In this abstract representation of the lifecycle (Mandi, 2020) we can see essentially two phases:

1. Detection
2. Response

When a disruption is detected (due to an adverse event) the system capabilities drop in their efficacy or performance level in some measurable amount or Ym. The system then recovers over a given time interval or Δτ due to some prebuilt resiliency or adaptability capability level (either partial or full).

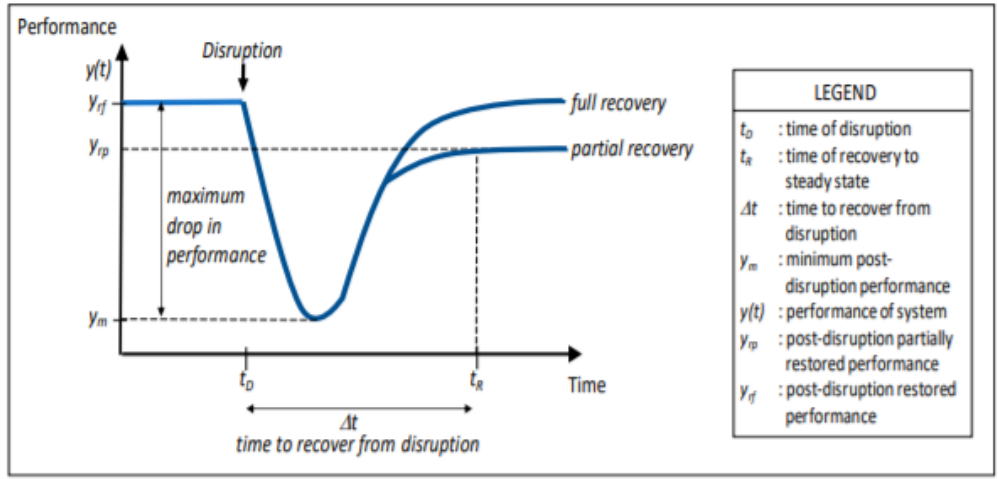

*Figure 2 –Resilience curve classically defined as a rebound function (Madni, 2020).*

As in the simplified phase diagram below (Figure 3) we can see that the difference with resiliency modeling is that there is no emphasis on failing over and failing back necessarily. The focus is on having the primary system continuing operations even while degraded in performance levels and then rebounding to normal operations. This represents a philosophically different design concept compared with Disaster Recovery, for example, and is more akin to High Availability and Fault Tolerant design approaches. Naturally, if a Disaster Recovery approach is taken to achieve a resiliency objective it should be engineered appropriately to maximize the resiliency objectives as modeled here. However, DR solutions are considered narrower in effectiveness than resiliency solutions when attempting to maintain uninterrupted service (Jurczak, 2016).

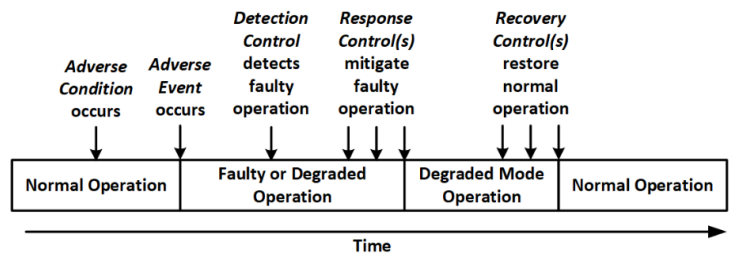

*Figure 3 - Fault Detection, Degraded Mode Operations, and Recovery Cycle Model (Firesmith, 2019)*

## 2.4   Resilience Operations Timeline

If we now convert this fundamental resiliency timeline into a more detailed operational view of how resiliency plays out in a dynamic sense, we see the model in the figure below provided by Ferris (2019). In this operational view there is a Threat, a Time vector, a set of actions or lack of actions, for a System of Interest (SoI). Resiliency is a topic of study in Information Security and Cyber-Physical systems and this explains the use of the term "threat" in this model. However, we can substitute this with the term "adversity", "fault", or "hazard" as we might see fit.

There are a few steps shown in Figure 4 from Ferris (2019):

P a g e 6 | 19

1. In this model there is time occurring prior to the event. During this time-period the system might be auto-observing itself or its environment trying to detect threats.
2. It might also be running self-diagnostics or conducting other preparation steps to take advantage of its non-threat state.
3. Eventually a threat will come into existence. Typically, there is lag time between this event and the time at which the system understands the event to exist.
4. In some cases, resistance to the threat may be effective in this early phase.
5. Potentially, resistance fails, and the resilience phase must protect the system.
6. Eventually, the system will then either regain its efficacy and functionality - or it will not - and it will then enter a failure state.

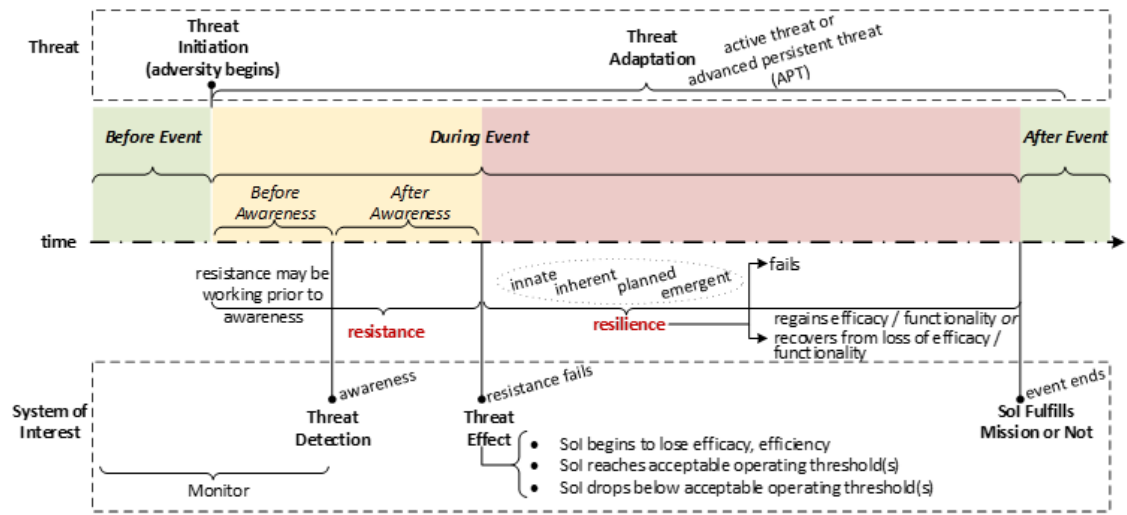

*Figure 4 - Operational View of a Resilience Timeline where SoI = System of Interest (Ferris, 2019)*

To summarize, the critical functions required when engineering a resiliency operational management process include primarily the same process steps seen in ITIL's Incident Management process:

- Monitor
- Detect
- Conduct triage
- Notify

The core difference here is the assumption that the system capability of resiliency should allow for continuation of service as opposed to an interruption and restoration of service which is the default model built into the ITIL process. The types of responses, also from Ferris, can be summarized as follows (these we be discussed in more detail below):

- Withstand
- Resist
- Change or Reconfigure
- Restart
- Fail-over (invoke redundancy)
- Fail-gracefully (tolerate)
- Fail-safe (revert to a safe condition)
- Recover (adapt, restore)

## 2.5 Failure Sources

Given this framework for the behavior of adversities and threats which drive system resiliencies against design thresholds it is important to understand the range of such events that can occur. We can begin this discussion with a relatively simple hierarchical breakdown of hazard types as seen in Figure 5 below. The ReSIST (Resilience for Survivability in Internet Systems) standard represented here provides a framework of environmental change classification to help provide some structure in thinking about a range of adversities a system might face from the natural world which may or may not be foreseen and those which might be short or long term (Meyer, 2019). This classification can be tailored to any problem domain and built out to finer grained needs to meet the requirements at hand.

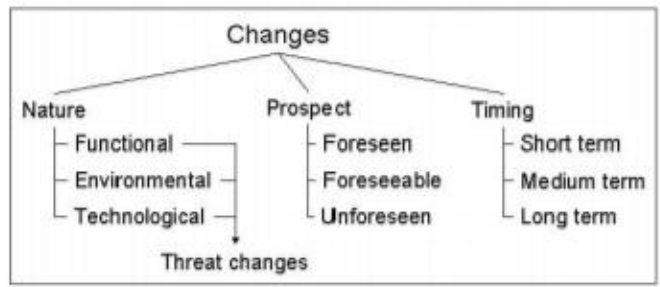

*Figure 5 -  ReSIST Classification of Changes (Meyer, 2013)*

Taking this one step further we can see in Figure 6 that if we expand the number and types of fault categories which might impact our system then the lattice of threats or adversities becomes a dense and puzzling matrix indeed.

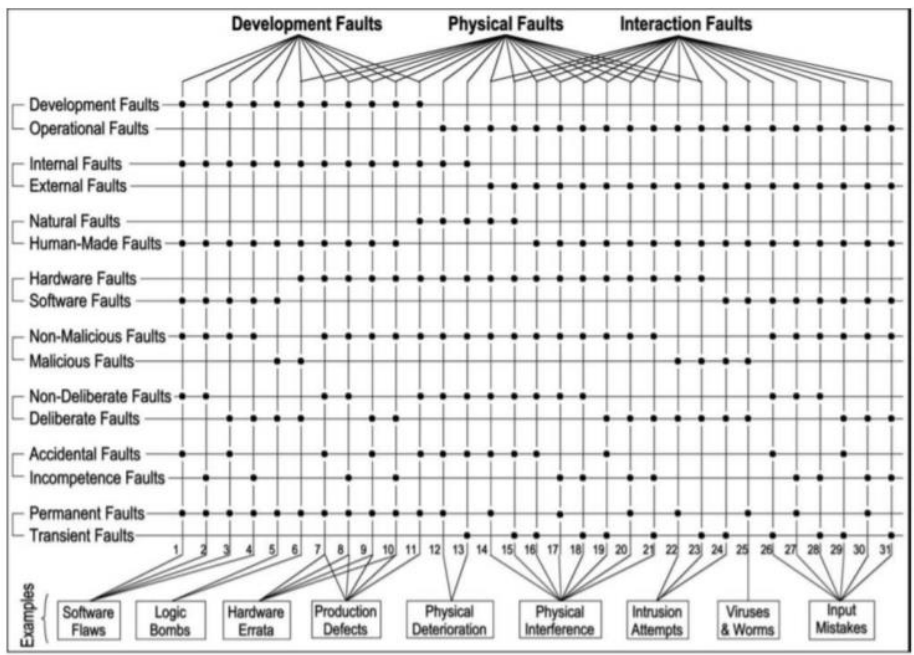

*Figure 6 - Fault Classes – A non-Exhaustive Universe of Event Types which a System must Adapt to Dynamically to achieve Resiliency (Meyer, 2013)*

The complexity of both the resiliency lifecycle above and the density of the potential universe of fault types shown here should be convincing that a remedy is required to this multi-threat condition. In the following sections we will provide approaches to that need.

# 3 Design for System Resiliency

Considering the above definitions and dynamics of Systems Resiliency we can turn to exploring a design response for such conditions. Because of the clear complexity of system resiliency requirements, the designs to combat the inherent challenges of resiliency solutions are not always straightforward. However, there are proven and cataloged methods, approaches, and design patterns in this domain. As most systems, by their very nature, bring with them the dual flavor of business benefit and operational risk, the designer is faced with a series of trade-offs. In the remaining sections we will discuss some general systems principles, details of resiliency factors, resiliency design patterns, failure mode analysis, and finally provide a few resiliency examples.

## 3.1 A Systems Theory

A good place to start when considering design is to discuss the nature of systems themselves. In an early treatment of this subject (Gall, 1977) a set of principles was proposed around systems and systems thinking. The first rule was that "everything is a system". Once that is accepted the designer can begin to think about certain key immutable aspects of how systems operate (a few choice ones are listed below) and then how to design for those operational aspects of systems as they cannot be avoided.

### 3.1.1 System Operations Rules

First, there are some core operational principles any system designer needs to understand prior to even starting on the design of a system:

1. Systems in general work poorly or not at all.
2. Some complex systems actually work.
3. Complex systems usually operate in failure mode.
4. A complex system can fail in an infinite number of ways. (If anything can go wrong, it will. *See Murphy's law*.)
5. The mode of failure of a complex system cannot ordinarily be predicted from its structure.
6. The larger the system, the greater the probability of unexpected failure.
7. Systems run better when designed to run downhill.

As a result of these summary rules we often see apparent breakdowns and even chaos around us in the modern world. Even though we fly millions of passenger miles a day or a year we still have plane crashes[1]. Why? Because we are running systems and systems follow the above rules. We also take for granted systems that work well and with limited friction. Yet it is those systems which required elegant design trade-offs to reach those levels of minimal entropy.

---

[1] See the recent tragic losses of two Boeing 737 Max aircraft. Root cause included overly complex override autopilot climb out software combined with lack of adequate pilot training on system changes.

### 3.1.2 System Design Rules

Thus, if systems are by their very nature error prone and inherently tend to fail then it would seem we should be able to counteract such "antics" through appropriate designs. Gall (1977) sets us up in our exploration of such design approaches and thinking with a few fundamental rules:

1. A simple system, designed from scratch, sometimes works.
2. A complex system that works is invariably found to have evolved from a simple system that works.
3. A complex system cannot be "made" to work. It either works or it does not.
4. The crucial variables (of design success) are discovered by accident.
5. Loose systems last longer and work better.

Interestingly, this guidance predates Agile methods by over two decades, yet it does encapsulate the same thinking to many degrees (i.e., minimal viable product, get it working and keep it working). This thinking also reflects the research of the time around loose coupling which was a relatively new design approach (at least in software) at the time but remains highly effective. The primary lesson here is that resiliency is a collective property baked into a system at the earliest stage of design. Retrofitting resiliency is either impossible or cost prohibitive so it should be an intentional part of the design mindset and process from the outset.

## 3.2 Resiliency and Costs

Each of these system drivers also lead us to risk analysis and the tradeoffs of design (probability analysis, consequence approximation). Such trade-offs force comparisons in optimizing resiliency vs. the costs of available mitigation strategies as shown in the diagram below (White, 2020). The important point is that a system engineering effort is not conducted in a business vacuum. There will always be a budget constraint as well to be factored into design decisions (see Figure 7).

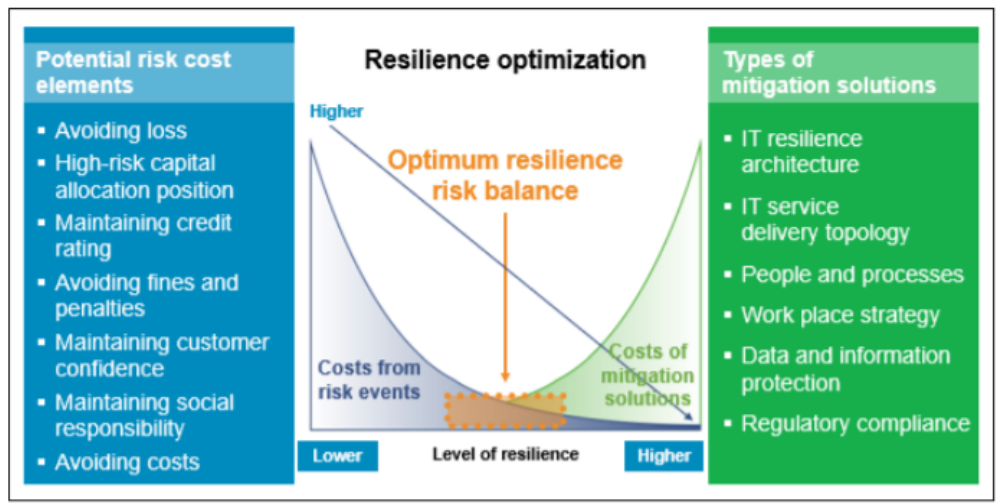

*Figure 7 - Resiliency optimization trade-offs (White, 2020)*

An example of this from the Bell System Engineering handbook focuses on "diminishing returns" (Harris, 1977). In this example, by adding a given number of phone circuits, call blocking would be reduced at certain call volume levels. However, at some point this reduction becomes very expense with ever smaller gains (i.e., from a probability of 0.01 down to 0.001). This implies that the optimization tradeoff steps in the design process is critical to arrive at what level of resiliency is affordable.

## 3.3 System Resiliency Factors

Designing for resiliency requires us to first peel back the onion on the factors briefly introduced above which influence resiliency in system outcomes. In general, we can think of these factors at a high level in the properties of deterrence, detection, delay, and response (Ed-daoui, 2018). There are also associated factors of system connectedness, system dependency, usability, and even economics such as the cost of a failure or the cost to repair. And as introduced above, the array of threats needs to be considered and matched against the system of interest (SoI). This brings us to the diagram below which maps out a dependency model for system resiliency (Figure 8).

As each box on the graph must be satisfied for a system to realize the full properties of being resilient (if applicable) or the risk assumed the requirements set is significant. This map provides a clear set of requirements classifications to be met through resilient design as we will now discuss in more specific detail.

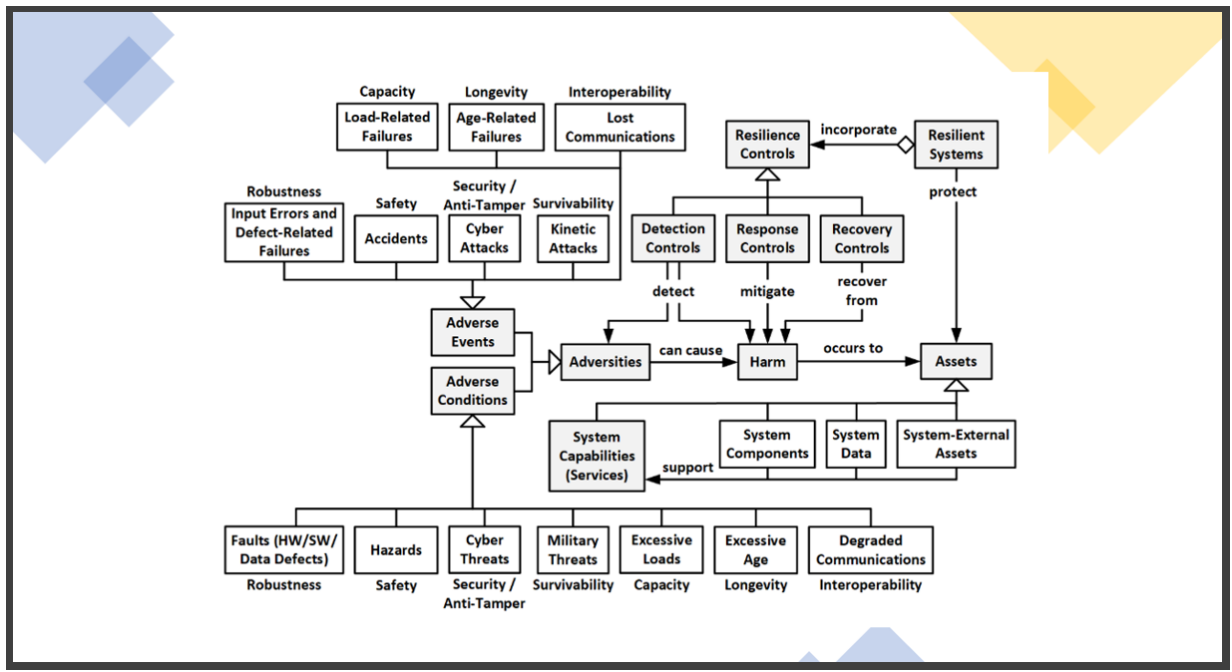

*Figure 8 - System Resiliency Dependency Model (Firesmith, 2019)*

## 3.4 Resiliency Design Patterns

Design-time consideration of resiliency requirements have been considered for decades and collected in engineering catalogs. In recent years these have been denoted as patterns following the work of Christopher Alexander (1979). Alexander's work in establishing patterns for architecture in part led Ward Cunningham to develop Agile methods utilizing techniques including pattern languages. Such pattern languages have now been extended throughout many disciplines and are a standard practice in representing reusable solution templates. This applies in our current discussion as follows:

*A resilience design pattern language provides the lexicon, syntax, and grammar to help articulate the abstractions of recurring resilient themes. The design patterns and the pattern language help systems engineers design solutions that provide resilience and systems that have the ability to be resilient.* (Ferris, 2019)

For example, a few specific resiliency patterns include:

- The absorption technique.
- The physical redundancy technique.
- The functional redundancy technique.
- The human in the loop technique.
- The distributed capacity technique.

Jackson (2016) recently published a catalog of such resiliency design patterns. These are summarized in the table below. In his original survey article Jackson also provides detailed source references for each sub-method so that the designer can easily dig into the next level of detail on the required resiliency solution approach as required. Naturally, for any given system not every resiliency design solution or pattern will be required simultaneously. However, understanding the full scope of available resiliency patterns, where they can be applied, and how, is critical.

| Resiliency Design Approach | Definition | Resiliency Design Sub-Methods |
|---|---|---|
| **1. Absorption** | The system should be capable of withstanding the design level disruption. | • **Margin** – The design level should be increased to allow for an increase in the disruption.<br>• **Hardening** – The system should be resistant to deformation.<br>• **Context spanning** – The system should be designed for both the maximum disruption level and the most likely disruption.<br>• **Limit degradation** – The absorption capability should not be allowed to degrade due to aging or poor maintenance. |
| **2. Restructuring** | The system should be capable of restructuring itself. | • **Authority escalation** – Authority to manage crises should escalate in accordance with the severity of the crisis.<br>• **Regroup** - The system should restructure itself after an encounter with a threat. |
| **3. Reparability** | The system should be capable of repairing itself. | • N/A |
| **4. Drift correction** | When approaching the boundary of resilience, the system should be able to avoid or perform corrective action; action can be taken against either real-time or latent threats. | • **Detection** – The system should be capable of detecting an approaching threat.<br>• **Corrective action** – The system should be capable of performing a corrective action following a detection.<br>• **Independent review** – The system should be capable of detecting faults that may result in a disruption later. |
| **5. Cross-scale interaction** | Every node of a system should be capable of communicating, cooperating, and collaborating with every other node. | • **Knowledge between nodes** – All nodes of the system should be capable of knowing what all the other nodes are doing.<br>• **Human monitoring** – Automated systems should understand the intent of the human operator.<br>• **Automated system monitoring** - The human should understand intent of the automated system.<br>• **Intent awareness** – All the nodes of a system should understand the intent of the other nodes.<br>• **Informed operator** - The human should be informed as to all aspects of an automated system.<br>• **Internode impediment** – There should be no administrative or technical obstacle to the interactions among elements of a system. |
| **6. Complexity Avoidance** | The system should not be more complex than necessary. | • **Reduce Variability** – The relationship between the elements of the system should be as stable as possible. |
| **7. Functional redundancy** | There should be two or more independent and physically different ways to perform a critical task. | • N/A |
| **8. Physical redundancy** | The system should possess two or more independent and identical legs to perform critical tasks. | • Nancy Leveson uses the term "*design redundancy*". |
| **9. Defense in depth** | The system should be capable of having two or more ways to address a single vulnerability. | • N/A |
| **10. Human in the loop** | There should always be human in the system when there is a need for human cognition. | • **Automated function** – It is preferable for humans to perform a function rather than automated systems when conditions are acceptable.<br>• **Reduce Human Error** – Standard strategies should be used to reduce human error.<br>• **Human in Control** – Humans should have final decision-making authority unless conditions preclude it. |
| **11. Loose Coupling** | The system should have the capability of limiting cascading failures by intentional delays at the nodes. | • **Containment** – The system will assure that failures cannot propagate from node to node. |
| **12. Modularity** | The functionality of a system should be distributed through various nodes of that system so that if a single node is damaged or destroyed, the remaining nodes will continue to function. | • N/A |
| **13. Neutral State** | Human agents should delay in taking action to make a more reasoned judgement as to what the best action might be. | • N/A |
| **14. Reduce Hidden Interactions** | Potentially harmful interactions between elements of the system should be reduced. | • N/A |

*Table 1 – Resiliency Design Patterns*

## 3.5 Resiliency and non-Functional Design Considerations

Beyond selecting appropriate resiliency design patterns there is a holistic problem to solve for. In specific, the architect must consider how the selection of specific design parameters and characteristics including non-functional requirements will influence overall system success and influence resiliency. As we have shown, resiliency is composed of numerous contributing factors. The architect's role is to satisfy each of those including the composite requirements of which some of the most critical are listed here (Bass, 1998):

- **Availability** – defined as the proportion of time the system is available for use. This is the first and most obvious capability of the system and an outcome of resiliency measured as below.

$$a = \frac{MTTF}{MTTF + MTTR}$$

- **Performance** – typical considered as system responsiveness, once the system is available. Performance requirements can be especially critical in a resiliency solution even when capabilities are degraded.

- **Security** – security requirements are critical under standard conditions but more so when the system is under duress. Resisting unauthorized usage attempts during denial of service attacks for example is a form or resiliency.

- **Reliability** – this is classically defined as the proportion of failure free operations over time (Musa, 1987) as shown here for software:

$$R(\tau) = e^{(-\lambda\tau)}$$

Where reliability $R$ is given through the negative asymptotic relationship between computer execution time $\tau$ and the failure rate $\lambda$. Reliability also has a direct relationship to Availability by converting reliability from a probability to a percentage simply multiplying by 100 as per below (Cusick, 2017):

$$R(\tau), \% = (e^{-\lambda\tau}) * 100\%$$

These measures drive resiliency design at a component and system level. Understanding the application of reliability to system design is critical to realizing resilient systems as discussed below.

## 3.6 Reliability Allocation and Resiliency

Once the architect has worked through understanding the functional needs of the application and has made the tradeoffs around the resiliency design patterns above an architecture emerges. These non-functional requirements will need to be specified and eventually verified. Further, reliability capabilities must be allocated to the components within the architecture to explicitly meet the first the overall system reliability objectives and then the overall resiliency needs of the solution. This means that for each component whether network, hardware, virtual server, software application, etc., the reliability, availability, and resiliency approach must be known and computed in relation to each other to tally up to the overall system reliability and comprehensive resiliency. This is where the linkage between reliability and resiliency is established.

The standard method for achieving the reliability understanding of the system at large is to layout the architectural components in a block diagram and assign their associated reliability ratings. Quantitatively, we then multiply the reliability rating of each component together to reach the collective reliability of the system (Musa, 1987). In the Reliability Block Diagram (RBD) below (Figure 9) this method is demonstrated and the computation for this approach is given below (Raza, 2019). This reliability calculation can then lead to an understanding of where the system resiliency capabilities will be found lacking.

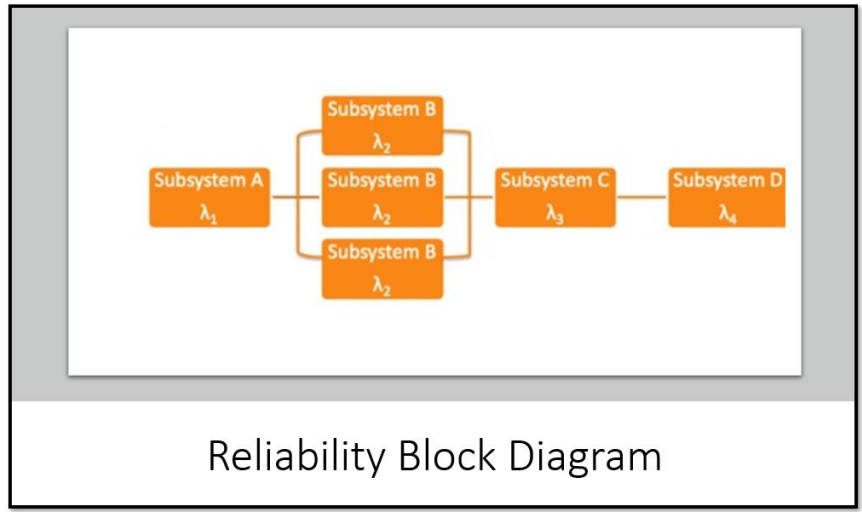

*Figure 9 - Computing reliability for distributed system resiliency (Raza, 2019)*

As an example, if computing for an N series-connected component architecture or an N parallel-connected array the approaches follow these standard formulas (Raza, 2019) as per Table 2:

| $$R(t) = \prod_{i=1}^{N} Ri(t)$$ $$A(t) = \prod_{i=1}^{N} Ai(t)$$ | $$R(t) = 1 - \prod_{i=1}^{N}(1 - Ri(t))$$ $$A(t) = 1 - \prod_{i=1}^{N}(1 - Ai(t))$$ |
|---|---|
| *Computation for N series-connected components. Note reliability drops as a product of the connected components.* | *Computation for N parallel-connected components. Note reliability improves in this configuration across nodes.* |

*Table 2 – Reliability Block Computations*

The key point of RBD analysis as demonstrated in these models is that for resiliency design simply adding more nodes **does not** necessarily increase reliability or resiliency. The designer must allocate reliability to components to derive the multiplicative reliability as given by the design and not assume that the architecture inherits higher reliability from the compositional units. As per Raza:

> *… two components with 99% availability connect in series to yield 98.01% availability. The converse is true for parallel combination model. If one component has 99%* *availability specifications, then two components combine in parallel to yield 99.99% availability …*

Thus, to be clear, resiliency is not gained only via distributed computing, parallelism, failover configurations, or Disaster Recover. Without additional design steps such as absorption, avoidance of complexity, reconfigurability, or other active methods classical high-availability techniques can also be overwhelmed at their threshold points whether they are engineered or accidental.

## 3.7 Failure Mode and Effect Analysis

Working from the component and architecture perspective is important in understanding reliability and resiliency. However, the number of design options and the volume of features for modern systems creates combinatorial complexity around potential faults which can be daunting to the analyst and designer. This calls for a methodology to conquer this complexity and systems engineering provides an answer.

The classic approach for understanding and managing failures is FMEA (Failure Mode and Effect Analysis). This discipline was established in the 1940s and has evolved since then especially in the aerospace industry (Snee, 2007). In Figure 10 below is a standard FMEA process diagram which can guide analysis around potential failure modes in a system and for determining their causes with the objective of putting specific controls in place around those causes.

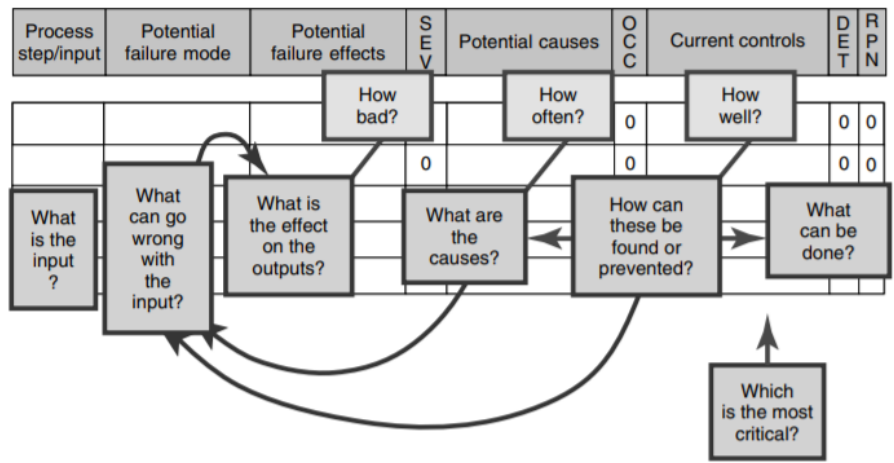

*Figure 10 - Failure Mode and Effect Analysis Chart (Snee, 2007)*

Essentially, during the design phase this process is applied to analyze and predict what type of threats, faults, and adversities the system might face. This also provides for the categorization of failure modes in terms of likelihood and impact. Next, beginning with the fundamental and long proven design patters for resiliency as presented above we demonstrate how we expect to manage these failure modes. In this way the system solution can predict, manage, and reduce the impact of the kinds of threats and adversities we have been discussing. The output of this analysis should then be hit against the solution architecture to revalidate the reliability and resiliency computations and scenario modeling to determine if in fact the designs will hold up in the face of the threat behaviors anticipated in the field.

## 3.8 System Resiliency Maturity and Metrics

Jackson (2016) also suggests that once a system is in place it should have a set of resiliency metrics. He argues that these metrics can proceed in a ladder of maturity as shown below. Such objective thinking and measurement on its face is a reasonable and productive path to focusing on the improvement or resiliency architectures, solutions, and operations.

- **Stage 1** – Existing system with no resiliency measurement.
- **Stage 2** – Resiliency principles at an initial level and applicable metrics follow from such improvements.
- **Stage 3** – Design for resiliency has been applied and specific measures are apparent and defined.
- **Stage 4** – System has encountered threat it was designed to encounter.

## 3.9 Resiliency Design Example

To help bring the concepts of resiliency and resiliency design to life an example is provided below. This resiliency design example is formatted in the style presented by Mandi (2020). This example is selected form the author's experience in work with architecture reviews at Bell Laboratories (Cusick, 1995).

**Telecom Switching - Load Shedding**: *Disruption*: (call overload on network)
- Modifiers: number of switches in network, number of calls per minute (time to drain traffic)

**Resilience Strategies Applied** (adaptive behaviors)
- Allow per switch traffic load to ramp up to 80% of pre-set threshold limit.
- At 80% limit begin out of band auto-signaling to peer switches to re-route traffic.
- Accomplish shedding of load to achieve balance below threshold or await human-in-loop.

# 4 Recommendations

Below are a set of recommendations consider when applying the concepts and approaches of resiliency to system solutioning problems.

## 4.1 Resiliency Strategy

1. IT organizations should develop and maintain a documented strategy around Systems and Operational Resiliency. This should include a policy statement and set of clear objectives.
2. Based on an analysis of the above definition of Systems Resiliency and the related methods to achieve resilient designs businesses need to develop specific plans to realize the recommendations as provided in this section and guided by the resiliency strategy.
3. Apply resiliency approaches as appropriate for the IT nature of the given IT environment.

## 4.2 Business Considerations

1. This document has focused almost exclusively on systems resiliency. Systems do support a particular business; however, an analysis of what resiliency means a particular busines is called for and a custom fit of the application of the use of System Resiliency methods to a given business environment may vary.
2. Organizational Resiliency is the companion piece to systems resiliency and has not been explored in this document. Organizational resiliency covers topics such as organizational, resources, staffing, facilities, and related processes to support Business Continuity. With the detailed definition and methodology around resiliency provided above it should be possible to apply these concepts to build out an approach to planning for organizational resiliency.

### 4.3 Methodology
1. Infuse development practices with essential systems resiliency thinking and design practices.
2. Train staff on the concepts, definitions, and methods of systems resiliency design.
3. Ensure that each system or application includes resiliency requirements.
4. Adopt resiliency design patterns (i.e., absorption, threshold management).
5. Quantify and specify target Availability in advance.
6. Allocate required reliability to meet specified Availability using RBD methodology.
7. Apply Failure Mode analysis.

### 4.4 Platform
1. Review and strengthen existing platform designs from a resiliency perspective.
2. Examine HA designs. Determine where failover configurations could be improved for resiliency.
3. Review recent failure trends and consider RCAs from a resiliency perspective.
4. Reconsider each infrastructure initiative from the point of view of a resiliency definition.
5. Analyze Cloud architecture availability requirements and the requisite resiliency response.
6. Double check platform architecture for single points of failure and compensate.
7. Explore use of Software Fault tolerance methods such as Process Pairs, Recovery Blocks, N-Version Programming (Alam, 2009).

## 5 Conclusions

We began this exploration by defining resiliency. We then looked at what it took to design for resiliency. We concluded by discussing the steps it might take to put this knowledge to work to improve the capabilities of IT systems. The purpose of this simply put is to achieve behaviors in applications and systems that are self-correcting in the face of adversity. It is the author's suggestion that by adopting a systems engineering orientation with a focus on resiliency design this capability envelope can be steadily pushed outward so that over time existing and new systems can achieve ever higher levels of self-resiliency to the benefit of customers, the business, and developers themselves.

## 6 About the Author

James Cusick is an IT leader with over 30 years of experience in Software Engineering, Software Reliability Engineering, IT Operations, Information Security, Process Engineering, and Project Management. James is currently Sr. Director IT Strategy and Operations with a global Information Services firm. James has held leadership roles with companies including Dell Services, Lucent's Bell Labs, and AT&T Laboratories. James is currently a Board Trustee at the Henry George School of Social Science where he researches topics in Innovation and Economics. James was also Adjunct Professor at Columbia University teaching Software Engineering. He has published widely in the above areas including two recent books on IT and Software Engineering. James is a graduate of both the University of California at Santa Barbara and Columbia University. James is a Member of the IEEE Computer Society and a PMP (Project Management Professional). Contact the author at j.cusick@computer.org.